\documentclass[sigconf,9pt]{acmart}
\AtBeginDocument{%
  \providecommand\BibTeX{{%
    Bib\TeX}}}

\usepackage[utf8]{inputenc}
\usepackage[bottom]{footmisc}
\usepackage{subcaption}
\usepackage{graphicx} 
\usepackage{algorithmic}
\usepackage{graphicx}
\usepackage{textcomp}
\usepackage{multirow}
\usepackage{enumitem}
\usepackage{xspace}
\usepackage{pifont}
\newcommand{\cmark}{\ding{51}} % checkmark
\newcommand{\xmark}{\ding{55}} % xmark
\usepackage{booktabs}
\usepackage{url}
\usepackage{caption}    %  \captionof

\usepackage{siunitx}

\usepackage{colortbl}
\newcommand{\MRN}[1]{{\color{black}#1}}
\newcommand{\spinge}[1]{{\color{cyan}[spinge: #1]}}
\usepackage{tikz}
\newcommand*\circled[1]{\tikz[baseline=(char.base)]{
            \node[shape=circle,fill,inner sep=0.5pt] (char) {\textcolor{white}{#1}};}}

\setlist[itemize]{noitemsep, topsep=0pt, leftmargin=*}
\begin{document}

%%
%% The "title" command has an optional parameter,
%% allowing the author to define a "short title" to be used in page headers.
\title{Cross-Layer Co-Designed In-Memory Hyperdimensional Computing Accelerator for Proteomics at the Edge \vspace{-5pt}}

\author{Md Mizanur Rahaman Nayan}
\affiliation{
  \institution{Georgia Institute of Technology}
  \country{}
}
\email{mnayan6@gatech.edu}

\author{Zheyu Li}
\affiliation{
  \institution{University of California San Diego}
  \country{}
}
\email{z5li@ucsd.edu}

\author{Flavio Ponzina}
\affiliation{
  \institution{San Diego State University}
  \country{}
}
\email{fponzina@sdsu.edu}

\author{Sumukh Pinge}
\affiliation{
  \institution{University of California San Diego}
  \country{}
}
\email{spinge@ucsd.edu}

\author{Tajana Rosing}
\affiliation{
  \institution{University of California San Diego}
  \country{}
}
\email{tajana@ucsd.edu}

\author{Azad J. Naeemi}
\affiliation{
  \institution{Georgia Institute of Technology}
  \country{}
}
\email{azad@gatech.edu}

%%
%% The "author" command and its associated commands are used to define
%% the authors and their affiliations.
%% Of note is the shared affiliation of the first two authors, and the
%% "authornote" and "authornotemark" commands
%% used to denote shared contribution to the research.

%%
%% By default, the full list of authors will be used in the page
%% headers. Often, this list is too long, and will overlap
%% other information printed in the page headers. This command allows
%% the author to define a more concise list
%% of authors' names for this purpose.
% \renewcommand{\shortauthors}{Trovato et al.}
%\newcommand{\Design}[0]{\textsc{EAGER}}

%%We eliminate the need for expensive re-clustering through a hardware-friendly incremental update mechanism while preserving nearly the same proteomics outcome."
%% The abstract is a short summary of the work to be presented in the
%% article.
\begin{abstract}
Database (DB) search and clustering are fundamental to data-analytics workloads such as mass spectrometry-driven proteomics. Current workflows rely on resource-intensive search and re-clustering, often offloaded to the cloud, raising privacy concerns for sensitive biological data. In this work, we enable efficient and privacy-preserving interaction on low-resource hardware platform through a cross-layer co-design approach. At the algorithm level, we propose a hardware-friendly data driven incremental update mechanism while preserving nearly the same proteomics outcome, eliminating costly repeated re-clustering during database updates. Performing cluster expansion directly during similarity search maps the algorithm naturally onto content-addressable memory (CAM), fusing distance computation and cluster expansion into one in-memory operation and eliminating dedicated distance units and data movement. This enables reduced footprint making a compact CAM-based accelerator feasible. We exploit bucket-wise search to scale and parallelize by mapping spectra buckets to CAM arrays to serve large dataset and maximize throughput, respectively. Bucket cache and main memory enables mapping datasets of different scales. Experimental results show a $\approx20\times$ algorithmic speedup over re-clustering with only $0.3\%$ additional error and $96\%$ DB search overlap with state-of-the-art(SOTA) methods. Bucket-wise parallelization enables $\approx100\times$ further acceleration over sequential search. The hardware implementation with $24.4MB$ SOT-MRAM CAM with $2.5MB$ bucket cache sustains $3.21\mu\text{J}$ per $1\text{K}$ queries and $263.8\mu\text{s}$ maximum latency on a $131GB$ human proteome dataset.

% Experimental results demonstrate a $20\times$ speedup over conventional full re-clustering with only a $0.3\%$ increase in clustering error. 

% DB search achieves $96\%$ overlap with state-of-the-art methods, validating result fidelity.

% With bucket-wise parallelization, the system attains up to $100\times$ additional speedup under sufficient CAM capacity.

% The hardware implementation uses a $24.4MB$ SOT-MRAM-based CAM, combined with a memory hierarchy consisting of a first-level cache and main memory, to support scalability for large-scale dataset achieves state-of-the-art search accuracy and clustering quality, while consuming $1.14\,\mu\text{J}$ per $1\text{K}$ queries and delivering $146\,\mu\text{s}$ average latency per search on a human proteome-scale dataset.
\end{abstract}

%Database search and clustering are fundamental components of many data analytics problems, including mass spectrometry driven proteomics.  Conventional full clustering and search algorithms have high resource usage and long latencies.\MRN{ We propose a lightweight incremental clustering method and a highly parallelizable DB search platform that leverages SOT-MRAM in-memory computing.} By leveraging mass-spectrometry insights, we employ bucket-wise parallelization and query scheduling to reduce latency. A one-time hardware initialization with pre-clustered proteomics data enables continuous DB search and local re-clustering, offering a more practical and efficient alternative to clustering from scratch. Heuristics from pre-clustered data guide incremental clustering, accelerating the process by $20\times$ with only a $0.3\%$ increase in clustering error. DB search results overlap by $96\%$ with state-of-the-art tools, validating search quality. The hardware leverages a $3T2MTJ$ SOT-MRAM based CAM at the $7nm$ node with a compute-in-memory design. 


%\ccsdesc[500]{xx~xx}

% \keywords{Proteomics, Hyperdimensional Computing, Compute in Memory, MRAM}

\settopmatter{printacmref=false, printccs=false, printfolios=false}
\renewcommand\footnotetextcopyrightpermission[1]{}
\maketitle

%----- Remember this -----
% I want to show two phases of interaction with hardware resources during MS-based proteomics. One is the initial clustering phase, followed by continuous database search and occasional clustering. The first phase requires very high resource usage and is needed only once, whereas the second phase is the regular case and is still performed using the same resources. This results in high resource usage, high energy consumption, and privacy concerns, as it often requires web-based services. We aim to replace this second phase with a resource-efficient system, HERP.

\vspace{-10pt}
\section{Introduction}
%Build the story as following:
% \begin{enumerate}
%     \item importance of Mass spectra search 
%     \item talk about whole clustering and user perspective. their frequency.
%     \item challenges and recent use of HD application
%     \item unaddressed challenges in the current HD based solution: huge data management, real time search, not suitable resource constraints env (energy and resources)
%     \item emerging tech like mram speciality and why it's a good tech to solve the problem compared to pcm, rram, feram.
%     \item our work to enable the application within resource constraint hardware with real time performance with less energy.
%     \item from user perspective: search and cluster expansion. that enables low resource execution without impacting performance.
%     \item highlight the contributions in architecture level and algorithm level.
% \end{enumerate}

Mass Spectrometry (MS) is used for many applications such as material discovery,  food safety, proteomics and clinical diagnostics, etc\cite {de2007mass, shuken2023introduction, aebersold2003mass, meissner2022emerging}. A key step in MS-based proteomics is searching through a very large database, where new variants are matched against large spectral libraries\cite{shuken2023introduction}. As a result, MS-based proteomics is very data-intensive. For instance, resources such as the MassIVE repository are approaching the petascale\cite{massive_repository, kim2014draft}. Searching across these massive datasets is extremely resource-intensive, with end-to-end runs often requiring many hours\cite{xu2023hyperspec}. 

Clustering is used to both improve the speed and the accuracy of search. Spectra are clustered based on similarity, leading to orders of magnitude more efficient hierarchical search\cite{wang2018mscrush, the2016maracluster, bittremieux2025large,xu2023hyperspec}. \MRN{However, tools such as MaRaCluster\cite{the2016maracluster}, msCRUSH\cite{wang2018mscrush}, and Falcon\cite{bittremieux2025large} run only on CPUs and  have not been sufficiently parallelized. This is crucial for large-scale dataset's real-time clustering and search performance \cite{fan2024specpcm}. ClusterSheep\cite{to2021clustersheep} introduces GPU accelerated clustering, but at the cost of clustering quality and search accuracy.} HyperSpec\cite{xu2023hyperspec}, a tool based on Hyperdimensional computing (HDC), offers GPU-based fast and high-quality clustering and DB search. HDC shows promise for encoding and analyzing mass spectra due to its inherent massive parallelization, as well as its efficient and accurate data compression, searching, and clustering capabilities~\cite{xu2023hyperspec,pinge2024spechd,SPECTRAFLUX,fan2024specpcm}. For context, HDC is a brain-inspired computing paradigm that encodes information into high-dimensional vectors (hypervectors, or HVs)~\cite{dewulf2024hyperdimensional,kleyko2023survey}. It relies on simple computational primitives—such as element-wise multiplication, addition, and bit shifting—that are well-suited for parallelization on various devices~\cite{nayan2025hydra, heddes2023torchhd,schlegel2022comparison}. However, accelerating HDC solutions using GPUs is limited by significant data movement and the need for a large pair-wise distance matrix, both of which negatively affect latency and energy consumption~\cite{fan2024specpcm}. 

Recent works have explored In-Memory Computing (IMC) approach using emerging non-volatile memories such as PCM and RRAM for mass spectrometry data analysis~\cite{fan2024specpcm, fan2024efficient}. This shows potential of inherently parallel distance calculation with no data transfer costs such as latency or energy~\cite{fan2024specpcm, li2022survey, zhang2022hd}. However, these devices come with major integration challenges. PCM faces the challenge of high error rate (10\%) and RRAM suffers from device variations\cite{fan2024specpcm, haensch2023compute}. Although HDC has high noise tolerance and resilience against bit errors~\cite{kleyko2022vector}, this comes at a cost of increased vector size that increases proportionally to the amount of noise to be tolerated\cite{thomas2021theoretical}. This results in large memory requirement. Another major problem is that they suffer from low endurance\cite{haensch2023compute}, which is critical during frequent updates on the database. 

All of the mentioned implementations suffer from expensive re-clustering during database updates and lacks scalability. Incremental clustering has been introduced very recently to avoid the re-clustering from scratch\cite{wang2026scalable}. However, this approach still requires server-class CPU due to expensive computation~\cite{wang2026scalable}. The reliance of existing solutions on heavy computation resources e.g., server-class CPUs and GPUs often forces users—such as researchers working with sensitive biological data—to depend on cloud-based services, thereby raising privacy concerns where the ideal solution demands the computation enabled on localized systems commonly available right next to mass spectra machines and ensures reliable, fast and accurate data analytics.

To this end, this work, HERP reduces the computation intensity first algorithmically by hardware friendly Hamming distance based incremental clustering guided by a data-driven similarity threshold unlike SOTA method which is heavily dependent on Falcon\cite{wang2026scalable}. We enable proposed cluster expansion algorithm directly in-memory during similarity search resulting in resource efficient hardware. CAM-based in-memory distance computation eliminates the need for separate distance computation units and significantly reduces data movement. Finally, We exploit bucket-wise search to scale and parallelize by mapping spectra buckets to CAM arrays to serve large dataset and maximize throughput. In summary, the main contributions can be broken down into three layers of abstraction: 
\begin{itemize}
    % \item \textbf{Algorithmic.} We leverage already available clustering results obtained from the large-scale database as a seed to avoid expensive computation at the target user's site. The integration of lightweight incremental clustering algorithm replaces full re-clustering, achieving faster execution while maintaining good clustering and search quality. 
    \item \textbf{Algorithm.} We develop a new efficient method of updating DB during search achieving faster execution on the proposed hardware while maintaining good clustering and search quality. The process is guided by a data-driven similarity threshold obtained from the pre-clustered data.
    \item \textbf{Architecture.} We introduce bucket-wise search mapped across CAM arrays to scale and parallelize. The proposed hardware supports CAM based in-memory distance computation with in-memory cluster expansion. Bucket caching technique is used for reducing latency while managing large dataset.
    \item \textbf{Technology.} We demonstrate the proposed hardware implementation with spin-orbit-torque magnetic random-access memory (SOT-MRAM) based 3T2MTJ CAM cell as the fundamental IMC unit at the $7nm$ CMOS technology node. 
    
\end{itemize}

\vspace{-5pt}
\section{Background and Related Works}
In this section we describe the key steps involved in mass spectrometry for proteomics,  followed by a description of HDC algorithm and its application in proteomics. Lastly, we discuss IMC-based clustering, search techniques and their challenges.
\begin{figure}
    \centering
    \vspace{-10pt}
    \includegraphics[width=\linewidth]{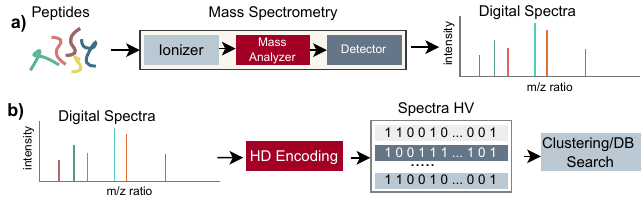}
    %\vspace{-10pt}
    \caption{Proteomics pipeline with HDC. a) Mass spectrometry transforms biological marker peptides into digital spectra b) HD encoding transforms the digital spectra into HVs.}
    \vspace{-10pt}
    \label{fig:MSFlow}
\end{figure}
\vspace{-10pt}
\subsection{\textbf{Mass Spectrometry and Proteomics}}
\hspace{10pt}\textbf{MS Pipeline:} In proteomics, biological samples are analyzed by Mass spectrometry to obtain a digitized spectrum (Fig. \ref{fig:MSFlow}(a). Peptide ions are generated by ionizer, separated by a mass analyzer according to mass-to-charge (m/z) ratio before detection~\cite{de2007mass, cravatt2007biological, aebersold2016mass}. The processed signal yields an intensity-versus-m/z spectrum that we encode as HVs for clustering and database search (Fig.\ref{fig:MSFlow}(b). 

\textbf{Clustering and DB Search:} These are the two primary tasks in proteomics. Clustering groups together spectra with similar characteristics and thus reduces the time needed for the search and increases its accuracy.  During DB search, the query spectra are matched to a spectral library. Matching candidates are filtered with a false discovery rate (FDR) unit to evaluate accuracy\cite{elias2007target}. Matched query provide already known information to understand query spectra while mismatch represents new variant. 

\textbf{Bucket Division:} During clustering, spectra are compared pairwise. Distance matrix is used to track the pairwise distance to find the most similar ones. The size of the matrix grows with spectra count in quadratic $O(n^2)$ complexity resulting in demand for massively sized memories  and excessively long search latencies. To avoid dense pairwise matrix spectrum comparison when clustering a large dataset, after pre-processing, spectra are sorted and assigned to a bucket based on their m/z value~\cite{bittremieux2025large, to2021clustersheep} according to the equation below,
\vspace{-3pt}
\begin{equation}
\text{bucket}_i = \left\lfloor \frac{(m/z_i - m_q) \times C_i}{d_c} \right\rfloor
\end{equation}
 where $C_i$ is the precursor charge, $m_q$ is charge mass 1.00794, $d_c$ is 1.0005079 which represents adjacent cluster distance and $m/z_i$ is associated with $i^{th}$ spectrum\cite{xu2023hyperspec}. In addition to reducing the search space, we further utilize this bucket to parallelize operations across CAM arrays during the proposed database search and clustering.

\vspace{-5pt}
\subsection{\textbf{HDC in Proteomics}}
\label{hdc-proteomics}
% Hyperdimensional computing emerged as an energy efficient and noise tolerant computing paradigm in which information is presented in a high dimensional space. Simple encoding techniques enable HDC to be adopted in resource constrained environments, and holographic information representation makes it robust against errors originating from device variation, noise in communication channels, and other faults. 
% HDC has already been used in MS Clustering and DB search for data compression, high quality clustering, and DB search quality~\cite{pinge2024spechd}. HDC also enables deploying MS Clustering and DB search in emerging memories like PCM, RRAM etc, to address errors originated from the device variations. To encode a spectrum generated after MS, ID-Level encoding scheme\cite{imani2017voicehd} is normally used\cite{fan2024specpcm, pinge2024spechd, xu2023hyperspec} where ID HV represents the peak m/z and level HV represents the peak intensity value. The two vectors are bounded by the \textit{XOR} operation. All HVs corresponding to the pairs exist in a spectra are bundled together to form the final HV for the spectra as follows:
\begin{figure}[b]
    \vspace{-8pt}
    \centering
    \includegraphics[width=\linewidth]{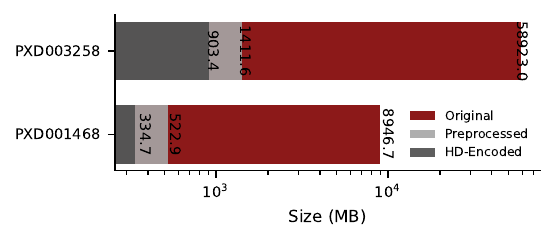}
    \vspace{-20pt}
    \caption{Dataset memory usage breakdown after preprocessing and HD encoding}
    \label{fig:sizeBreakdown}
    \vspace{-10pt}
\end{figure}

HDC is an energy-efficient, noise-tolerant paradigm where information is represented by HVs in high-dimensional space. Its simple encoding schemes make it suitable for resource-constrained environments, while holographic representation ensures robustness against device variation, channel noise, and bit errors. HDC has been successfully applied to MS clustering and DB search, enabling data compression, high-quality clustering, and accurate search results~\cite{pinge2024spechd}. Fig.~\ref{fig:sizeBreakdown} illustrates the compression achieved by HD encoding followed by raw spectra pre-processing.  HDC maps naturally onto emerging memories such as MRAM, PCM and RRAM, mitigating errors due to device variability. For spectra encoding, the commonly used ID-Level scheme~\cite{imani2017voicehd} represents the peak m/z with an ID HV and the peak intensity with a Level HV; the two are combined via \textit{XOR}, and all resulting HVs are bundled to form the final spectrum HV~\cite{fan2024specpcm, pinge2024spechd, xu2023hyperspec} where Majority(.) function transforms the HVs into binary HVs, P represents the pairs of intensity and m/z are the values of the spectra:

\vspace{-8pt}
\begin{equation}
        \mathbf{h} = \text{Majority}\left( \sum_{(i,j) \in \mathbb{P}} I_i \oplus L_j \right)
\end{equation}

\begin{figure*}
    \vspace{-3pt}
    \centering
    \includegraphics[width=0.9\textwidth]{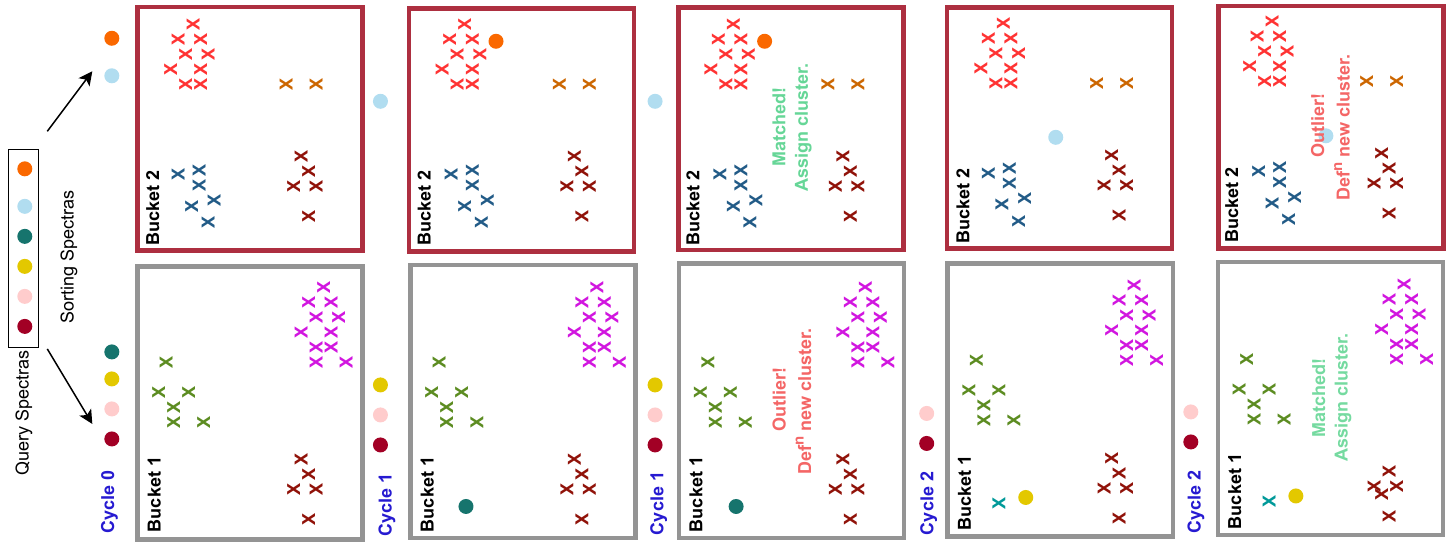}
    \vspace{-5pt}
    \caption{Simplified walkthrough example of the proposed DB search and simplified cluster expansion. From user end if a query is matched against a clustered bucket, it is assigned to the cluster. In case of a mismatch, a new cluster is formed.}
    \label{fig:example}
    \vspace{-20pt}
\end{figure*}
\vspace{-10pt}
\subsection{\textbf{IMC in MS Clustering and DB Search}}
Clustering and DB search both require a spectrum from an MS experiment to compare against a collection of spectra which is time-consuming and computationally expensive. Prior efforts have attempted to tackle this problem through techniques like hashing, approximate nearest neighbor search, and efficient dot product/similarity kernels \cite{bittremieux2025large, wang2018mscrush, arab2023semisupervised}, but their effectiveness is often limited by high-precision floating-point arithmetic. HDC clustering tools like HyperSpec\cite{xu2023hyperspec}, SpecHD\cite{pinge2024spechd} and DB search tools such as  HyperOMS\cite{kang2023accelerating}, RapidOMS\cite{pinge2024rapidoms} show that it is possible to get state of the art accuracy at high efficiency and parallelism using only binary operations. A recent study shows that although HD-powered clustering and DB search are beneficial, a major bottleneck is distance calculation\cite{fan2024specpcm}. The problem is severe when the dataset is large, which involves large-scale matrix computations leading to significant data movement, especially when a dataset exceeds the GPU's onboard memory capacity. 

Compute-in-memory-based systems using PCM address this challenge by reducing data movement and distance computation time due to parallel search\cite{fan2024specpcm, fan2024efficient}. However, they lack memory hierarchy to support dataset of different scale. Additionally, PCM 2T2R cell suffers from high error rate results in 4 write verify cycle and require higher HV dimension to withstand errors. ADC and DAC footprint occupy $47\%$ of chip area resulting in a large capacity reduction~\cite{fan2024specpcm}. Moreover, performing clustering from scratch every single time is slow and resource intensive. It can be avoided by initializing the system with pre-clustered centroids. 
\vspace{0pt}
% We therefore approach the problem from the user's perspective who generates spectra locally,  performs search on existing DB under resource constraints, where full clustering from scratch is uncommon; DB search is the frequent case, and new cluster heads are formed only when a mismatched query does not fit any existing cluster. With this work, we present a solution that integrates hardware–software co-design and leverages the SOT-CAM device along with their compute-in-memory capabilities.

\vspace{-10pt}
\section{Methodology and System Workflow}

This section presents the proposed methodology with a simplified walkthrough example to illustrate the proposed algorithm, followed by a description of the proposed hardware architecture and mapping of the algorithm. Finally, we describe the cell and array level functionalities of the CAM unit, which forms the core of the proposed hardware.
\vspace{-10pt}
\subsection{\textbf{Proposed Algorithm}} Fig.\ref{fig:example} illustrates the cycle-wise flow through a toy example consisting of two buckets. Each bucket contains its own clusters, represented by consensus spectra. These bucket-wise clusters and their corresponding consensus spectra are collected from the initially clustered spectra generated by SOTA clustering tools. The objective is to leverage this pre-clustered data for user-end applications, where new spectra are continuously searched and clusters are updated when necessary. \MRN{The processes involved during execution of} the example is split into three stages:

\textbf{Bucket Loading and Query Sorting.} Query spectra are staged for search against consensus spectra representing bucket clusters in \textit{\textbf{Cycle 0}}. After pre-processing, the query spectra are sorted by their mass-to-charge (m/z) ratio to determine the appropriate bucket. Once the bucket ID is assigned, the spectra are queued bucket-wise to enable sequential searches across buckets.

\textbf{Performing DB Search.} One query spectrum from each bucket queue is searched against the corresponding bucket clusters. Two outcomes are possible: \circled{1} the query spectrum matches an existing cluster, or \circled{2} it is an outlier, i.e., it belongs to a cluster that does not yet exist within the bucket. In this case, a new cluster is defined. In \textit{\textbf{Cycle 1}}, the query in Bucket 1 is an outlier, while the query in Bucket 2 matches an existing cluster. Similarly, in \textit{\textbf{Cycle 2}}, Bucket 1 has a match with the newly defined cluster, whereas the query in Bucket 2 does not match and is thus considered an outlier, leading to the creation of a new cluster in the next cycle.  

\textbf{Cluster Expansion.} The matched spectrum is assigned to the corresponding cluster ID where mismatched spectrum is used to define a new cluster instead of re-clustering the entire bucket. This significantly reduces execution time at the cost of negligible cluster quality loss. The decision of whether a spectrum is a match or an outlier is determined using a heuristic derived from initial clustering, where the minimum distance between the query and cluster spectra is compared against a dynamic threshold. 

\textbf{Thresholding Criterion.} 
After initial clustering, for each cluster $C$ we compute the Hamming distances $D_C$ between its consensus spectrum and assigned spectra, with mean $\mu_C$ and standard deviation $\sigma_C$. A query is a match if its nearest-consensus distance is below:
\begin{equation}
T_C = \mu_C + K_C\,\sigma_C .
\end{equation}
Rather than fixing $K_C$, we fix a single global coverage target $p$ (we use $p = 0.95$) and set $K_C = \bigl(Q_p(D_C) - \mu_C\bigr)/\sigma_C$, where $Q_p(\cdot)$ is the empirical $p$-th percentile. This makes $T_C = Q_p(D_C)$ by construction, so a fixed fraction $p$ of each cluster's own spectra fall within its radius; $K_C$ varies per cluster precisely because it absorbs the skew and heavy tails of real proteomics distance distributions. Storing $T_C$ in $\mu{+}K\sigma$ form lets the hardware keep compact per-cluster statistics. For sparse clusters ($|D_C| < n_{\min}$) the percentile is unreliable, so $K_C$ falls back to $K_{\mathrm{pool}}$ estimated from distances pooled across sparse clusters. Statistics are computed once on the seed data and recomputed only when the fraction of newly formed clusters exceeds a set threshold which is an inexpensive pass over stored distances but insignificant compared to re-clustering. Independently of the choice of $p$, Chebyshev's inequality guarantees intra-cluster coverage of at least $1 - 1/K_C^2$ without any distributional assumption, providing a distribution-free floor (e.g., $K_C = 3, 5$ give $\geq 88.9\%, 96\%$).

\label{hardware}
\vspace{-10pt}
\subsection{\textbf{HERP Hardware Architecture}}
\begin{figure}[t]
    \centering
    \vspace{-15pt}
    \includegraphics[width=\linewidth]{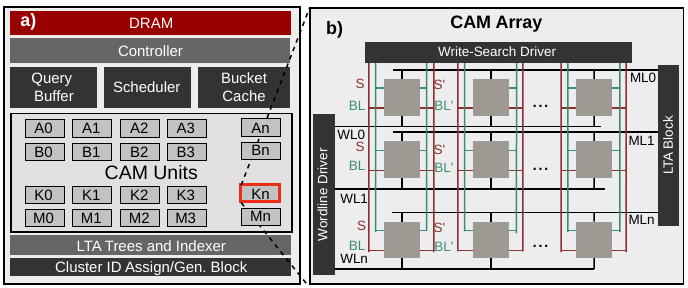}
    \caption{a) Proposed Hardware architecture b) SOT-CAM array, Core unit for similarity search.}
    \label{fig:architecture}
    \vspace{-15pt}
\end{figure}

Fig. \ref{fig:architecture} shows the proposed architecture overview. The Query buffer stores preprocessed spectra to be searched, after encoding into HVs. CAM units, Fig. \ref{fig:architecture}(b) store the consensus HVs representing clusters of the buckets. The scheduler keeps track of the buckets available in the CAM units and is also responsible for making the decision to evict a bucket from the CAM units at the time of an unavailable bucket demanded by query HVs. In that scenario, the bucket cache is looked up to check the availability of the requested bucket. If the bucket is missing, the controller generates a signal to request main memory for the bucket. The scheduler also sorts the spectra and forwards them to the corresponding FIFO buffer. From the FIFO buffer at each cycle, one query HV is dispatched to be searched across the CAMs of the corresponding bucket, which generates distances between the consensus HVs and the query HV. The Loser Takes All (LTA) tree shared across the CAMs is used to find the lowest distance. This distance is compared against the threshold to decide whether the spectra belongs to an existing cluster or a new cluster definition is needed. If there is a match then the cluster ID is generated from the index tracking of the LTA tree. For outliers that require a new cluster definition, a new ID is generated and assigned to the HV, and it is added to the CAM block representing the bucket in the next update. Two challenges arise when the dataset is large: \circled{1} HV size or the number of consensus spectra of a bucket can be too large to reside in a single CAM array which is  $128\times128$, \circled{2} the number of buckets can be too large to fit in the available CAM blocks. The issues are addressed as follows: \\
\hspace{10pt}\textbf{\circled{1} CAM array assignment.} CAM columns represent elements of a HV, and the rows represent different HVs. Multiple CAM blocks are used to map all the elements of a HV. Currents representing the distance from each CAM block are accumulated to represent distances between the query HV and the consensus HVs. 
\begin{figure}[b]
    \vspace{-10pt}
    \centering
    \includegraphics[width=1.0\linewidth]{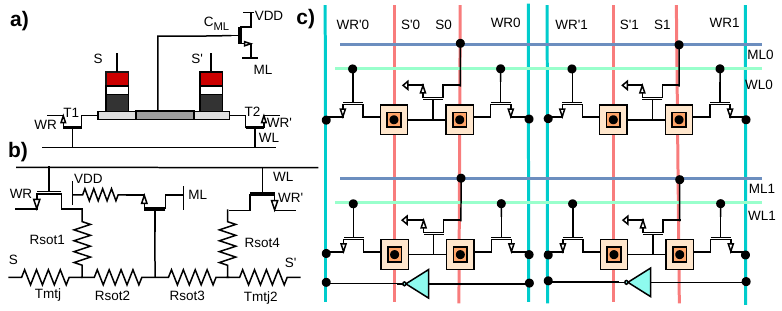}
    \caption{\MRN{a) 3T2MTJ SOT-CAM cell b) equivalent circuit of the cell c) 2x2 CAM array circuit.}}
    \vspace{-15pt}
    \label{fig:cellCktArr}
\end{figure}

% \begin{figure}
% \vspace{-5pt}
%     \centering
%     \includegraphics[width=0.9\linewidth]{figures/executionFlow.pdf}
%     \caption{Operational phases of the proposed system. }
%     \label{threephases}
%     \vspace{-5pt}
% \end{figure}
\textbf{\circled{2} Memory hierarchy for scalability.} It is theoretically impossible to accommodate all spectra in the CAM units simultaneously due to the large number of buckets. We addresses this limitation by allowing spectra to be searched independently across buckets. Thus, only the buckets demanded by the query spectra need to be available at a given time. Initially, smaller buckets are prioritized to maximize the number of buckets resident in the CAM unit. During the search, queries are sorted and organized according to the currently available buckets. As demand increases, additional buckets are brought into the CAM units by evicting less frequently used (LFU) buckets, while minimizing eviction overhead given the varying bucket sizes. We use bucket cache to reduce the latency caused by data transfers from main memory.
\vspace{-5pt}

\vspace{-5pt}
\subsection{\textbf{SOT-CAM as Fundamental Computing Unit}}

% PCM, RRAM, FeRAM, and MRAM are the major emerging non-volatile memory technologies. For industry adoption, candidates must meet key benchmarking metrics, including latency, energy, cell area, error rate, endurance, retention, and process maturity. Each device has unique characteristics that make it suitable for specific applications. Among them, SOT-MRAM stands out, offering cell density higher than SRAM, read latency below 1ns, write and search energy in the pJ range, error rates of $10^{-6}$, and endurance exceeding $10^{13}$. In comparison, PCM suffers from higher write latencies ($\sim$10ns), large write error rates ($\sim10\%$), and limited endurance ($10^9$). Process maturity further favors SOT-MRAM, as wafer-level fabrication has already been demonstrated, whereas FeRAM and others still face challenges such as device variability and high write voltages~\cite{yasin2024extremely}. The recent demonstration of a $64$Gb MRAM chip further establishes MRAM as a leading candidate among emerging NVMs~\cite{hatsuda202530}. Overall, SOT-MRAM shows clear superiority across the benchmarking metrics.
CAM array is primarily responsible for in memory search which is crucial for both DB search and incremental clustering. SOT-MRAM based (SOT-CAM) cells drive the low energy consumption and latency of the proposed hardware at the technology level. The array design and mapping of the search on the array are as follows:

\textbf{3T2MTJ SOT-CAM Cell.} Fig.~\ref{fig:cellCktArr}(a, b) illustrates the CAM cell \MRN{and the corresponding equivalent electrical circuit,} where the voltage at node $C_{ML}$ is high (low) when there is a mismatch (match) between the stored value and the search bit. The node $C_{ML}$ controls the NMOS device connected to the match line (ML), which is shared by all cells of a row in the CAM array (Fig.\ref{fig:cellCktArr}(c)). Note that complementary values are stored in the two MTJs, and complementary search voltages are applied on the search lines(SL) to reduce the error rate. The voltage division between the two MTJ's generates a high or low voltage at the $C_{ML}$ node. During the write operation, the word line (WL) is activated and the bit line is connected to $WR$ and $WR'$, which inject current through the U shaped SOT layer to align the MTJs spin state according to the applied bit line values.

\begin{figure}[t]
    \centering
    \vspace{-15pt}
    \includegraphics[width=\linewidth]{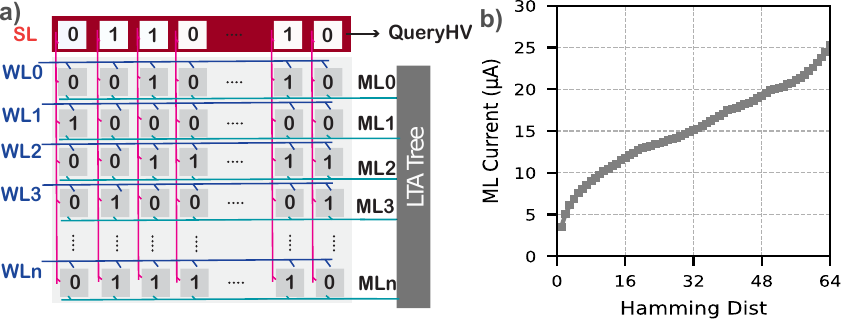}
    \caption{\MRN{a) Mapping search on CAM array. Matchline current represents the distance between query and the stored HV b) Current vs Hamming distance relationship considering parasitic of the circuit after using voltage scaling.}}
    \vspace{-5pt}
    \label{fig:mappingScaling}
    \vspace{-10pt}
\end{figure}

\MRN{\textbf{DB Search Mapping and Scaling.}} During the DB search, the Query HV is loaded into the SL and a similarity search is performed against all HVs within the array (Fig.\ref{fig:mappingScaling}(a)). The currents from the cells connected to the ML are summed, and the resulting current reflects the Hamming distance between the stored vector and the query. An LTA block is then used to identify the smallest current. \MRN{To address parasitic non ideality during scaling, we have used search voltage scaling to linearize current-distance relationship (Fig.\ref{fig:mappingScaling}(b) as presented in the prior work\cite{nayan2025hydra}.}

\vspace{-10pt}
\section{Experimental Evaluation}
\vspace{-5pt}
\MRN{This section describes the HERP implementation and evaluates its search and clustering quality, along with end to end latency improvements from algorithmic, architectural, and SOT-MRAM device innovations compared to SOTA methods. It concludes with an system evaluation and CAM technologies trade-off analysis.}
% \begin{figure}
% \vspace{-10pt}
%     \centering
%     \includegraphics[width=\linewidth]{figures/figure_venn_oms_no_single.pdf}
%     \caption{DB Search Identification Comparisons: Venn diagrams that depict the overlap of identified unique peptides using consensus spectra generated by HyperSpec, HERP-0.9, and HERP-0.7.   Identified peptides from HERP are highly overlapped with the results generated by HyperSpec baseline. }
%     \label{oms_venn}
% \end{figure}

\begin{figure}[t]
    \centering
    \vspace{-5pt}
    \includegraphics[width=0.8\linewidth]{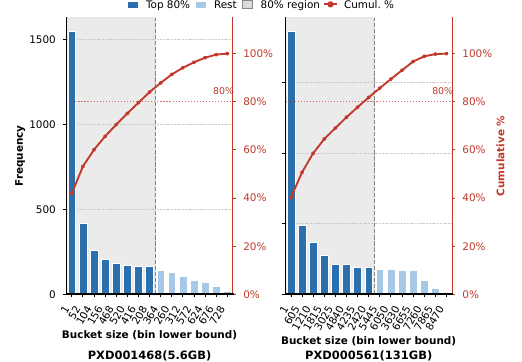}
    \vspace{-0pt}
    \caption{Distribution of clusters across buckets from the two pre-clustered dataset.}
    \vspace{-15pt}
    \label{fig:frequency_dist_data}
\end{figure}

% \begin{figure*}
%     \centering
%     \includegraphics[width=0.8\textwidth]{figures/pareto_side_by_side (1).pdf}
%     \caption{Distribution of clusters across buckets from two pre-clustered dataset. This demonstrates the opportunity of efficient bucket allocation on CAM units and caching to reduce bucket miss rate.}
%     \label{fig:placeholder}
% \end{figure*}

\begin{figure}[b]
    \centering
    \vspace{-8pt}
    \includegraphics[width=0.8\linewidth]
    {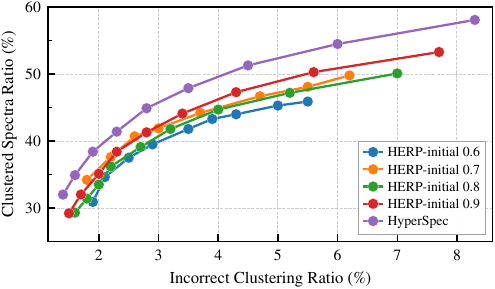}
    \vspace{-3pt}
    \caption{Clustering quality of HERP and HyperSpec\cite{xu2023hyperspec}}
    \vspace{-15pt}
    \label{quality_initial_ratio}
\end{figure}

% \vspace{-0pt}

% \begin{figure*}[t]
% \vspace{-5pt}
% \centering

% \begin{minipage}{0.48\textwidth}
%     \centering
%     \includegraphics[width=\linewidth]{figures/plot_quality_Initial_ratio.pdf}
%     \caption{Clustering Quality Comparisons: clustered spectra ratio vs incorrect clustering ratio.}
%     \label{quality_initial_ratio}
% \end{minipage}
% \hfill
% \begin{minipage}{0.48\textwidth}
%     \centering
%     \includegraphics[width=\linewidth]{figures/upset_charge.pdf}
%     \caption{Consensus UpSet plot showing overlap and unique identifications between HyperSpec\cite{xu2023hyperspec} and HERP.}
%     \label{oms_upset}
% \end{minipage}

% \vspace{-15pt}
% \end{figure*}
\vspace{-10pt}
\subsection{\textbf{Experimental Setup}}
\textbf{Dataset.}
% We have considered two dataset of diferent size. PXD001468\cite{chick2015mass}, PXD000561\cite{kim2014draft} which belong to kindney cell and human proteome cell type. Their size are about 5.6GB and 131GB, respectively. 
We evaluate two datasets of markedly different scales: PXD001468~\cite{chick2015mass} and PXD000561~\cite{kim2014draft}, corresponding to kidney-cell and human-proteome experiments, with sizes of approximately 5.6~GB and 131~GB, respectively. Fig. \ref{fig:frequency_dist_data} depicts the distribution of buckets in the two datasets by bucket size, where bucket size represents the number of clusters in each bucket. The Pareto chart highlights that 80\% of the bucket sizes fall within half of the maximum bucket size. This insight is useful for bucket allocation in CAM and caching to optimize latency. 

% However, accessing larger buckets incurs a higher penalty due to their size. Therefore, both the bucket access pattern and bucket size must be balanced during CAM allocation. 

% \textbf{Performance Metric.} Cluster spectra ratio, which assesses the clustering capability by keeping the incorrect clustering ratio fixed is used as clustering quality metric. We have compared the number of total identified peptides using the proposed method given the fixed FDR rate with those identified by other tools for DB search quality. For latency, we use average latency to search a bucket for a query. For energy, we measure the energy from the CAM cell for being main source of consumption. 

\textbf{Runtime Baseline Tools:} For end-to-end runtime evaluation, we compare HERP against SOTA CPU-based clustering tools MaRaCluster~\cite{the2016maracluster}, msCRUSH~\cite{wang2018mscrush}, Falcon~\cite{bittremieux2025large}, and the GPU-based HDC method HyperSpec~\cite{xu2023hyperspec}, which shows substantially faster clustering while maintaining SOTA accuracy. For conventional hardware baselines we use a server-grade Intel Xeon Gold 6548Y+ CPU with 128 logical cores and an NVIDIA L40S GPU with 48GB of memory.

\textbf{Hardware Specifications:} 
For hardware implementation of HERP using SOT-CAM, we have used the experimentally validated model for the SOT layer and MTJ \cite{narla2022design}. The MTJs have a diameter of $45\text{nm}$ and an oxide thickness ($t_{ox}$) of $2\text{nm}$, resulting in resistances of $1.25\text{M}\Omega$ in the parallel state and $3.44\text{M}\Omega$ in the anti-parallel state. A $3.3\text{nm}$ thick AuPt layer is used as the SOT channel, with the thickness optimized to minimize write energy based on the spin drift-diffusion model~\cite{zhu2018highly}. For the access transistor and the ML drive transistor we use NMOS from ASAP $7\text{nm}$ PDK. For CAM operation, the search voltage applied on the search line is set to $1\text{V}$ and the write voltage used on the bit lines is set to $0.8\text{V}$. We design a $128 \times 128$ SOT-CAM array and perform search and write operations to evaluate latency, power, and energy consumption using HSPICE. For fair comparison, the SPICE simulations also account for interconnect parasitic extracted from the physical layout. After characterization of the array and other peripherals like LTA tree and WL driver, we have used an in-house simulator for system evaluation. The system capacity of storing and searching across is $100K$ consensus HVs. Each HV is of $2048$ bits. We have set the HV dimension for all the datasets since it offers a good balance between performance and accuracy\cite{xu2023hyperspec, pinge2024spechd}. The system has CAM unit of size $24.4MB$ made of SOT-CAM and $2.5MB$ bucket cache made of SRAM. Each array has dedicated write driver and bit line driver units (Fig.\ref{fig:architecture}(b)) that help to parallelize the HV loading and search. 
\begin{figure}[t]
    \vspace{-5pt}
    \centering
    \includegraphics[width=0.8\linewidth]{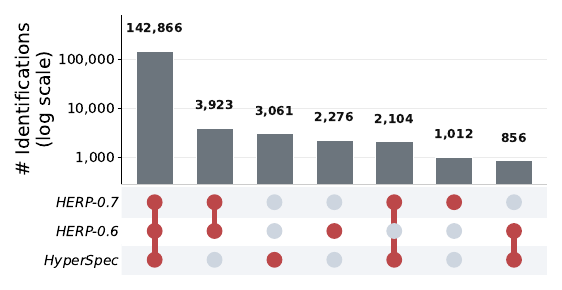}
    \caption{Consensus UpSet plot showing overlap and unique identifications between HyperSpec\cite{xu2023hyperspec} and HERP.}
    \vspace{-5pt}
    \label{oms_upset}
\end{figure}
% \begin{figure}[t]
%     \centering
%     \vspace{-5pt}
%     \includegraphics[width=0.7\linewidth]{figures/latency_bucket.pdf}
%     \caption{Average bucket search latency for different requested bucket location.}
%     \label{fig:bucket_latency}
%     \vspace{-10pt}
% \end{figure}
\vspace{-10pt}
\subsection{\textbf{Clustering and Search Quality}}
\hspace{10pt}\textbf{Clustering Quality.}
We evaluate the quality of HERP cluster expansion in Fig.~\ref{quality_initial_ratio}. A higher clustered spectra ratio at a lower incorrect clustering ratio reflects better clustering quality. Our approach begins by clustering a subset of the dataset, followed by incremental clustering of the remaining spectra through the proposed method. At clustered spectra ratio of 40\%, the HyperSpec baseline yields an incorrect clustering ratio of 2.5\%, while HERP-initial 0.6 (40\% of the dataset clustered via expansion) achieves 2.8\% demonstrating HERP’s clustering quality comparable to HyperSpec.
%, with a modest reduction in quality when using fewer initial data.
% \begin{figure*}[t]
% \vspace{-5pt}
% \centering

% \begin{minipage}{0.48\textwidth}
%     \centering
%     \vspace{-5pt}
%     \includegraphics[width=\linewidth]{figures/speeupComparision.pdf}
%     \caption{\MRN{ Speedup comparsion of HERP during incremental clustering over re-clustering.}}
%     \vspace{-5pt}
%     \label{fig:xUPincClustering}
% \end{minipage}
% \hfill
% \begin{minipage}{0.48\textwidth}
%     \vspace{-10pt}
%     \centering
%     \includegraphics[width=0.8\linewidth]{figures/speedupBreakdown.pdf}
%     \caption{Speedup Breakdown of HERP system over baseline.}
%     \vspace{-5pt}
%     \label{fig:ablationStudy}
% \end{minipage}

% \vspace{-5pt}
% \end{figure*}

\begin{figure}[t]
    \centering
    \vspace{-5pt}
    \includegraphics[width=0.8\linewidth]{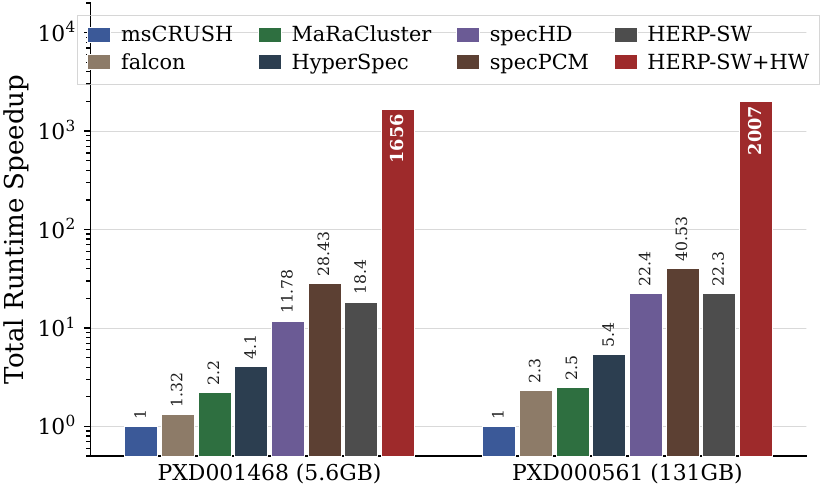}
    \caption{\MRN{ Speedup comparsion of HERP during incremental clustering over re-clustering.}}
    \vspace{-15pt}
    \label{fig:xUPincClustering}
\end{figure}

\textbf{DB Search Quality.}
Clustered datasets are primarily used for downstream DB search to identify peptide sequences. We compared DB search accuracy between the HyperSpec baseline and HERP, controlling the clustered spectra ratio to 40\% and FDR at 1\%. Fig.~\ref{oms_upset} illustrates the overlap of unique peptide identifications obtained from consensus spectra clustered by HyperSpec and HERP. The upset plot shows that HERP has 142866 identical overlap ($\approx96\%$ of found by HyperSpec) with the HyperSpec baseline. Since HyperSpec has already been shown to match or exceed the identification accuracy of prior CPU-based, non-HDC clustering pipelines~\cite{the2016maracluster},~\cite{wang2018mscrush},~\cite{bittremieux2025large}, this high overlap suggests that HERP preserves SOTA downstream DB search fidelity while focusing on further improving efficiency.

\textbf{Robustness.} We varied the initial seed size from 0.6 to 0.9 to demonstrate the proposed algorithmic robustness against the internal distribution shift, where 0.6 and 0.9 means 40\% and 10\% of the data is clustered by our algorithm respectively. Fig.\ref{quality_initial_ratio} illustrates that HERP results are very consistent in terms of clustered spectra ratio vs incorrect clustering ratio for different initial seed size implying the robustness of the algorithm against internal distribution shift. Note that, HERP algorithm is intended for within-distribution data expansion. If a seed dataset from one tissue type (e.g., kidney cells) is used to search spectra from a markedly different source, it would generate many sparse clusters with few assigned candidates. However, this is unlikely in practice, as seed and query data typically come from the same experimental context.

% \begin{table}[t]
% \vspace{-5pt}
% \centering
% \caption{\MRN{Comparison of CAM Technologies}}
% \setlength{\tabcolsep}{3pt} % reduce column spacing
% \resizebox{\columnwidth}{!}{%
% \begin{tabular}{
%     l
%     S[table-format=1.5]
%     S[table-format=1.1]
%     S[table-format=1.2]
% }
% \toprule
% \rowcolor{gray!30}
% \textbf{Parameter (CAM Cell)} & \textbf{SOT-MRAM} & \textbf{CMOS\cite{ni2019ferroelectric}} & \textbf{PCM\cite{li20131}} \\
% \midrule
% \rowcolor{green!30}
% Cell Structure & {3T2MTJ} & {16T} & {2T2R} \\
% Cell Area (\si{\micro\meter\squared}) & 0.0583 & 1.12 & 0.41 \\
% Search Energy per Bit (fJ) & 0.714 & 1.0 & {0.64} \\
% Search Latency (ns) & 0.485 & 0.75 & 1.9 \\
% Operating Voltage (V) & 0.8 & 1.0 & 2.5 \\
% Write Latency (ns) & 2 & 1 & 10 \\
% Write Energy per Bit (fJ) & 278 & 4.8 & 4500 \\
% Endurance & {\num{e13}\cite{nguyen2024recent}} & \text{Inf.} & {\num{e7}\cite{haensch2023compute}} \\
% Technology (nm) & 7 & 45 & 90 \\
% Write Verify Cycle & \xmark & \xmark & 4\cite{fan2024specpcm} \\
% Non-Volatile & \cmark & \xmark & \cmark \\
% \bottomrule
% \end{tabular}
% }
% \label{tab:cam_comparison}
% \vspace{-5pt}
% \end{table}

\vspace{-10pt}
\subsection{\textbf{Latency and Energy Profiling}}
\label{Latency_profiling}
\vspace{0pt}
In the proposed method, compute heavy initial clustering is avoided which takes around 3min 12s for kidney cell and 24min for human draft proteome in HyperSpec tool on GPU where other clustering tools like GLEAMS\cite{bittremieux2025large}, MaRaCluster\cite{the2016maracluster}, Falcon\cite{bittremieux2025large} require more than 2hr\cite{xu2023hyperspec}. HERP system stores bucket wise consensus spectra HVs in the main memory and then loads on the CAM units based on demand. For initial loading of the considered system under experiment, write energy is $56.93\mu J$ for 100K spectra with bucket count of 124 for the human genome draft proteome. Latency of loading(write) is 16ns which leverages parallel write in the arrays.  
\begin{figure}[t]
    \vspace{-15pt}
    \centering
    \includegraphics[width=\linewidth]{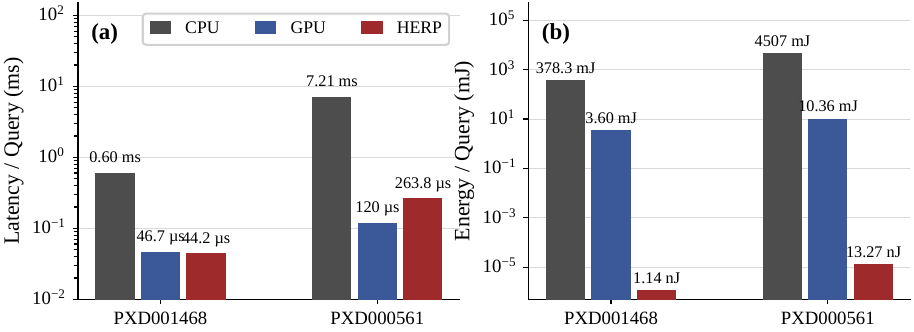}
    \caption{Hardware level comparison with CPU and GPU.}
    \vspace{-10pt}
    \label{fig:ablationStudy}
\end{figure}
\hspace{10pt}\textbf{Clustering:} The proposed approach uses incremental clustering instead of full bucket re-clustering unlike SOTA tools which brings significant speedup over existing tools. Fig.\ref{fig:xUPincClustering} shows the relative speedup gain during clustering. HDC powered near storage processing baselines like SpecHD\cite{pinge2024spechd} and SpecPCM\cite{fan2024specpcm} offers hardware level gain where HERP algorithmic advantage alone match the gain. Additionally, bucket wise parallelization and SOT-CAM based implementation achieves orders of magnitude clustering speedup.

\textbf{DB Search:} Search energy per query is dependent on the average bucket size that determines the search space. We have found average search energy per query is 0.27$nJ$ with average bucket size of 186 and max 1.14$nJ$ for the largest bucket of size 780 for PXD001468 (small) dataset. For PXD000561 (large), average search energy per query is 3.21$nJ$ with average bucket size of 2196 and max 13.27$nJ$ for the largest bucket of size 9075. The search latency in a bucket depends on the location of the requested bucket where latency is lowest when bucket is already in CAM unit and highest when is in main memory. Additionally, bucket size also affect latency if located in cache or in main memory as longer memory transaction is required for larger bucket. We have considered first level cache and average bucket size during the latency measurement. For PXD001468, latency per query is 3.75$\mu s$ and 44.19$\mu s$ considering the requested bucket in the cache and in the memory respectively. For PXD000561, they are 22.34$\mu s$ and 263.8$\mu s$ respectively.

\textbf{Hardware Trade-offs and Efficiency:} We have compared HERP system with CPU and GPU by running the same incremental algorithm with bucket parallelization enabled as per capacity. Fig.\ref{fig:ablationStudy} illustrates that the evaluated HERP system achieves comparable search latency per query with CAM unit capacity of only 24.4MB and cache size of 2.5MB while reducing search energy per query by orders of magnitude. This is directly driven by massive parallel in memory distance computation in SOT-CAM and simplified memory hierarchy. In contrast, CPU and GPU implementations require expensive XOR, popcount, and argmin kernels, alongside frequent back-and-forth data movement. These insights directly highlight the critical energy and resource efficiency benefits HERP delivers for enabling the edge-computing environments. 

\vspace{-13pt}
\subsection{\textbf{System Evaluation}}
\vspace{-3pt}
\hspace{10pt}\textbf{Mapping large scale dataset.} Section.\ref{Latency_profiling} discusses mapping and profiling of the two dataset of different scale on same hardware with 100K HVs CAM capacity which justifies the mapping efficiency of the proposed architecture. High compression ratio from pre-processing and encoding reduces (Section.\ref{hdc-proteomics}) the memory requirement to a great extent enabling mapping large scale data e.g., 131GB Human genome dataset on the proposed hardware which sits beside the mass spectrometry machine. The processed encoded HVs are stored in local computer main memory serving as main memory of the proposed hardware as well. 

\textbf{Overhead.} Bringing the distance computation in memory comes at some cost. We use 3T2MTJ SOT-CAM cell as a fundamental computing unit of size 0.05832$um^2$ where conventional SOT-MRAM cell requires 2T1MTJ of size 0.0322$um^2$ at 7nm technology node. The CAM unit occupies around $11.22~mm^2$. Additionally, LTA tree is used to detect the most similar one and to keep track of the index. HERP uses LTA trees of $\log_2(n)$ stage and shares across CAM arrays but still has footprint of 0.2081$mm^2$. Moreover, bucket cache of size $2.5MB$ is used occupying $1.57mm^2$ area for SRAM. Despite this overhead, HERP lowers energy consumption and latency by reducing both computational workload and data movement compared to SOTA tools\cite{fan2024specpcm, xu2023hyperspec, wang2018mscrush, the2016maracluster}.

\vspace{-10pt}
\subsection{Design Trade-offs in CAM Technologies}
The proposed cross layer co-designed platform is orthogonal to the choice of CAM technology. The design benefits from a CAM technology with higher endurance because during search often bucket miss happens, and CAM arrays need to be rewritten with requested HVs. Additionally, since the system is intended for deployment at the user end to address privacy concerns, non-volatility plays a crucial role in reducing overall energy consumption. We have chosen the SOT-CAM option in this work because of its high endurance, low error rate, long retention, CMOS compatibility, and process maturity. However, the hardware can be implemented using other technologies considering specific design goals. Table. \ref{tab:cam_comparison} highlights CAM implementations reported in the SOTA works to highlight the various trade-offs across technology options. Although SRAM offers the highest endurance, it suffers from the large cell area due to 10T cell and volatile nature. PCM and ReRAM suffer from low endurance. Beside SOT-MRAM based CAM, FeFET shows potential for relatively low write heavy datasets as recent work reports major advancements in endurance at high density\cite{chen2025novel}. 
\vspace{-0pt}
\begin{table}[htb]
\centering
\vspace{-10pt}
\caption{SOTA CAM Technologies}
\label{tab:cam_comparison}
\setlength{\tabcolsep}{4pt}
\renewcommand{\arraystretch}{0.9}
\resizebox{\columnwidth}{!}{%
\begin{tabular}{@{}lccccc@{}}
\toprule
\textbf{Metric} & \textbf{CMOS}\,\cite{9373942} & \textbf{FeFET}\,\cite{9372119} & \textbf{ReRAM}\,\cite{10745585} & \textbf{PCM}\,\cite{li20131} & \textbf{SOT-MRAM} \\ \midrule
Cell type & 10T & 2 FeFET & 2T-2R & 2T-2R & 3T-2MTJ \\
Endurance\cite{haensch2023compute, yu2022semiconductor} & $>10^{16}$ & $10^9$ & $10^5$ & $10^7$ & $10^{15}$ \\
Search Latency (ns) & 0.75 & 0.341 & 0.351 & 2.5 & 0.485 \\
Operating Voltage (V) & 1.0 & 4.0 & 1.0 & 2.5 & 1.0 \\
Search energy (fJ/bit) & 1.02 & 0.35 & 0.55 & 0.64 & 0.714 \\
Cell area ($\mu$m$^2$) & 2.66 & 0.15 & 0.41 & 0.41 & 0.058 \\
Node (nm) & 28 & 45 & 90 & 90 & 7 \\
Non-Volatile & \xmark & \cmark & \cmark & \cmark & \cmark \\
\bottomrule
\end{tabular}%
}
\vspace{-8pt}
\end{table}

\vspace{-10pt}
\section{Conclusion}
\vspace{-2pt}
DB search and clustering are central to proteomics workloads, requiring resource and energy-efficient execution at the edge for ensuring privacy. To eliminate reliance on server-class CPUs/GPUs via cloud, we present HERP, a cross-layer co-designed solution that algorithmically lowers computational intensity to enable execution on resource efficient hardware, as well as offer custom architecture with memory hierarchy for energy efficient CAM based solution. The proposed algorithmic shifts and bucket wise parallelization enables speedup while in memory based solution offers high energy efficiency. Additionally, memory hierarchy and main memory sharing with local host enables serving the clustering and DB search for practical workloads. 
\vspace{-5pt}
\begin{acks}
This work was supported in part by CoCoSys and PRISM, two centers in JUMP 2.0, a Semiconductor Research Corporation (SRC) program sponsored by DARPA.
\end{acks}
\vspace{-0pt}

%%
%% The next two lines define the bibliography style to be used, and
\bibliographystyle{ACM-Reference-Format}
\bibliography{ref}
\end{document}